\documentclass[cameraready]{Interspeech}

\usepackage{amssymb,amsfonts}
\usepackage{booktabs}
\usepackage{multirow}
\usepackage{url}
\usepackage{threeparttable}
\title{KFC-KWS: Keyframe Fusion with CTC for User-Defined Keyword Spotting}

\author[affiliation={1}]{Jin}{Li}
\author[affiliation={2}, correspondingauthor]{Wenbin}{Jiang}
\author[affiliation={1}]{Ji}{Hu}

\address{
    $^1$ School of Electronics and Information Engineering, Hangzhou Dianzi University, Hangzhou, China \\
    $^2$ School of Communication Engineering, Hangzhou Dianzi University, Hangzhou, China
}

\email{\{jinli, wbjiang, huji\}@hdu.edu.cn}


\usepackage{comment}

\begin{document}

\maketitle

\begin{abstract}
User-defined keyword spotting (KWS) enables personalized voice interaction by detecting user-specified keywords. A key challenge in this task is distinguishing target keywords from phonetically confusable alternatives. To address this challenge, we propose KFC-KWS, a multimodal framework that leverages connectionist temporal classification (CTC)-guided keyframe selection. Specifically, we exploit the peaky posterior distributions of CTC to identify high-confidence phoneme frames, enabling precise alignment across audio, phoneme, and text modalities. These keyframes are then fused with full-utterance representations through cross-attention to capture both local discriminative cues and global contextual information. On LibriPhrase, KFC-KWS achieves the best-balanced performance (98.73\% AUC) and substantially outperforms advanced baselines on the challenging hard subset (97.65\% AUC and 7.75\% EER), demonstrating its effectiveness in discriminating between highly confusable keywords.
\end{abstract}

\section{Introduction}

\begin{figure*}
    \centering
    \includegraphics[width=0.95\linewidth]{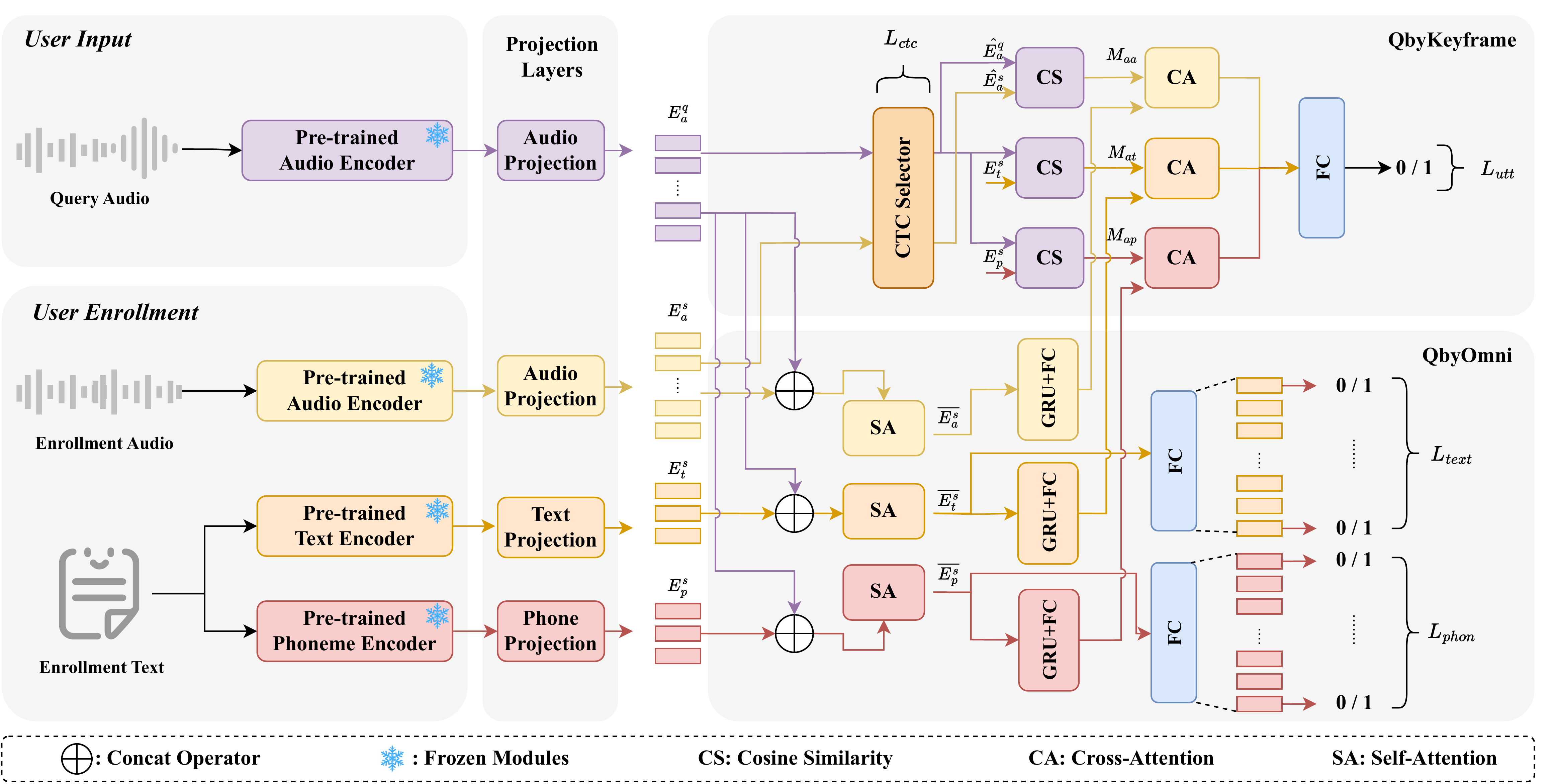}
    \caption{Overall architecture of the proposed method, KFC-KWS. The enrollment audio and text are converted into audio, text, and phoneme embeddings using different pre-trained encoders. The same pre-trained audio encoder processes the user input audio to obtain query audio embeddings. These query embeddings interact with each enrollment embedding through two branches, QbyOmni and QbyKeyframe, which are finally merged to output a confidence score.}
    \label{fig:pipeline}
\end{figure*}
Keyword Spotting (KWS) aims to detect whether a keyword appears in a speech segment and is essential for voice-interactive devices. Popular KWS methods train models on large collections of predefined keyword samples, thus supporting only fixed vocabularies \cite{zhu2024temporal,sarawgi2024streaming,saidutta2023wake,higuchi2020stacked}. To enable open-vocabulary detection, KWS systems adopt CTC \cite{wang2023wekws} or transducer frameworks, decoding keywords through beam search. However, these approaches often rely on costly training pipelines and remain unsuitable for efficient on-device adaptation to new user-defined keywords.

User-defined KWS (UD-KWS) addresses this limitation by allowing users to enroll keywords through text or audio \cite{jung2023metric,yang2022catt,ozay2024joint,shin2022learning,nishu2024flexible,zhang2023u2,liu2021rnn,gurugubelli2024comparative}. Text-based enrollment typically extracts phoneme features and matches them with audio embeddings \cite{yang2023keyword,liu2022neural,xi2022text}, providing stable performance but degrading under strong accents or noisy conditions. Audio-based enrollment, also known as query-by-example \cite{wang2024query,huang2021query,zhan2021stage}, better adapts to user-specific pronunciation and is commonly implemented using metric or meta-learning frameworks \cite{reuter2023multilingual,chen2018investigation,kao2023efficiency}. 

Recent studies further combine text and audio enrollment in multimodal frameworks to improve robustness. CLAD \cite{xi2024contrastive}, MM-KWS \cite{ai2024mm}, and PLCL \cite{kewei2024phoneme} align semantic, phonetic, and acoustic representations and achieve nearly perfect accuracy on easy keywords. However, reducing false activations caused by phonetically confusable keywords remains a core challenge in practical deployment, especially when keywords differ by only one or two phonemes. A key observation motivating our work is that such confusable pairs typically differ at only a few phoneme positions, while full-utterance matching methods treat all frames equally, diluting subtle but critical distinctions.

Models trained with CTC exhibit a characteristic ``peaky'' behavior~\cite{graves2006connectionist}, concentrating posterior mass on a small number of phoneme-aligned frames. These high-confidence frames carry the most discriminative phonetic information and provide natural anchors for cross-modal alignment. Building on this insight, we propose KFC-KWS, a multimodal framework that integrates CTC-guided keyframe matching with full-utterance feature fusion, as illustrated in Fig.~\ref{fig:pipeline}. Unlike prior methods that rely on uniform frame-level attention \cite{ai2024mm} or external phoneme memory banks \cite{kewei2024phoneme}, KFC-KWS directly leverages the CTC posterior as a zero-cost frame-importance indicator, enabling phoneme-aligned matching without additional modules. To further enhance robustness, we apply random modality masking during training as a lightweight regularization strategy.

Our approach differs from recent UD-KWS methods in several aspects. CLAD~\cite{xi2024contrastive} and MM-KWS~\cite{ai2024mm} perform full-sequence matching across modalities, treating all frames equally regardless of phonetic salience. PLCL~\cite{kewei2024phoneme} introduces a phoneme memory bank to enhance phoneme-level representations but requires additional external components, while AdaKWS~\cite{navon2024open} employs adaptive fusion without explicit phoneme-level alignment. In contrast, KFC-KWS uses the CTC posterior directly as a frame selector and fuses keyframe-level and utterance-level representations, balancing fine-grained discrimination and global context. 

Overall, the contributions of this work are as follows:
\begin{itemize}
    \item We introduce a CTC-guided keyframe selection strategy that exploits peaky posterior distributions to extract phoneme-aligned keyframes, enabling fine-grained cross-modal matching among audio, phoneme, and text modalities.
    \item On the LibriPhrase benchmark, KFC-KWS achieves the best balanced performance (98.73\% AUC) and substantially improves performance on the challenging hard subset, demonstrating its effectiveness in detecting confusable keywords.
\end{itemize}

\section{Proposed Method}
\label{sec:mtd}
\subsection{Model architecture}
For enrollment audio and text, we extract initial embeddings using
pre-trained encoders. The enrollment text is processed by both a G2P
converter (for phoneme features)~\cite{g2pE2019} and a text encoder (for semantic
features). The query and enrollment audio share the same encoder to
ensure aligned feature spaces.

Each modality-specific embedding is first passed through its
corresponding projection layer, transforming it into a unified feature
space. Specifically, all modality features are linearly projected to a
shared dimension, followed by the addition of positional encodings to
preserve temporal structure and modality embeddings to indicate modality
type. We denote the resulting representations as follows: the query
audio feature $E_a^q$, the enrolled audio feature $E_a^s$, the phoneme
feature $E_p^s$, and the semantic feature $E_t^s$, all in
$\mathbb{R}^{T \times D}$, where $T$ depends on the input sequence
length and $D$ is the unified feature dimension.

The final embedding for each modality is obtained by summing the
projected feature with its corresponding positional and modality
encodings. This shared representation format enables different
modalities to follow the same processing pipeline, ensuring that
features from heterogeneous sources are aligned in the same latent
space and facilitating cross-modal matching. The above feature
representations are then fed into two parallel branches:
keyframe-based querying (QbyKeyframe) and complete feature-based
querying (QbyOmni).

\textbf{QbyOmni.} In this branch, the full query audio features are
matched cross-modally with different enrolled modality features.
Specifically, we first concatenate $E_a^q$ with each of $E_a^s$,
$E_p^s$, and $E_t^s$ along the temporal dimension, and apply a
self-attention mechanism to obtain the enhanced representations
$\overline{E_m^s}$, defined as in Equation \ref{eq:qbyc}:
\begin{equation}
\label{eq:qbyc}
\overline{E_m^s} = \text{Self-Attention}( E_a^q \oplus E_m^s ), \quad m \in \{a, p, t\},
\end{equation}
Subsequently, a gated recurrent unit (GRU) followed by a fully
connected (FC) layer is used to map the complete feature for each
modality into a fixed-dimensional sequence $F_m^c \in \mathbb{R}^{T \times D}$,
as defined in Equation \ref{eq:cf}:
\begin{equation}
\label{eq:cf}
F_m^c = \text{FC}(\text{GRU}(\overline{E_m^s})),
\end{equation}

\begin{figure}
    \centering
    \includegraphics[width=1\linewidth]{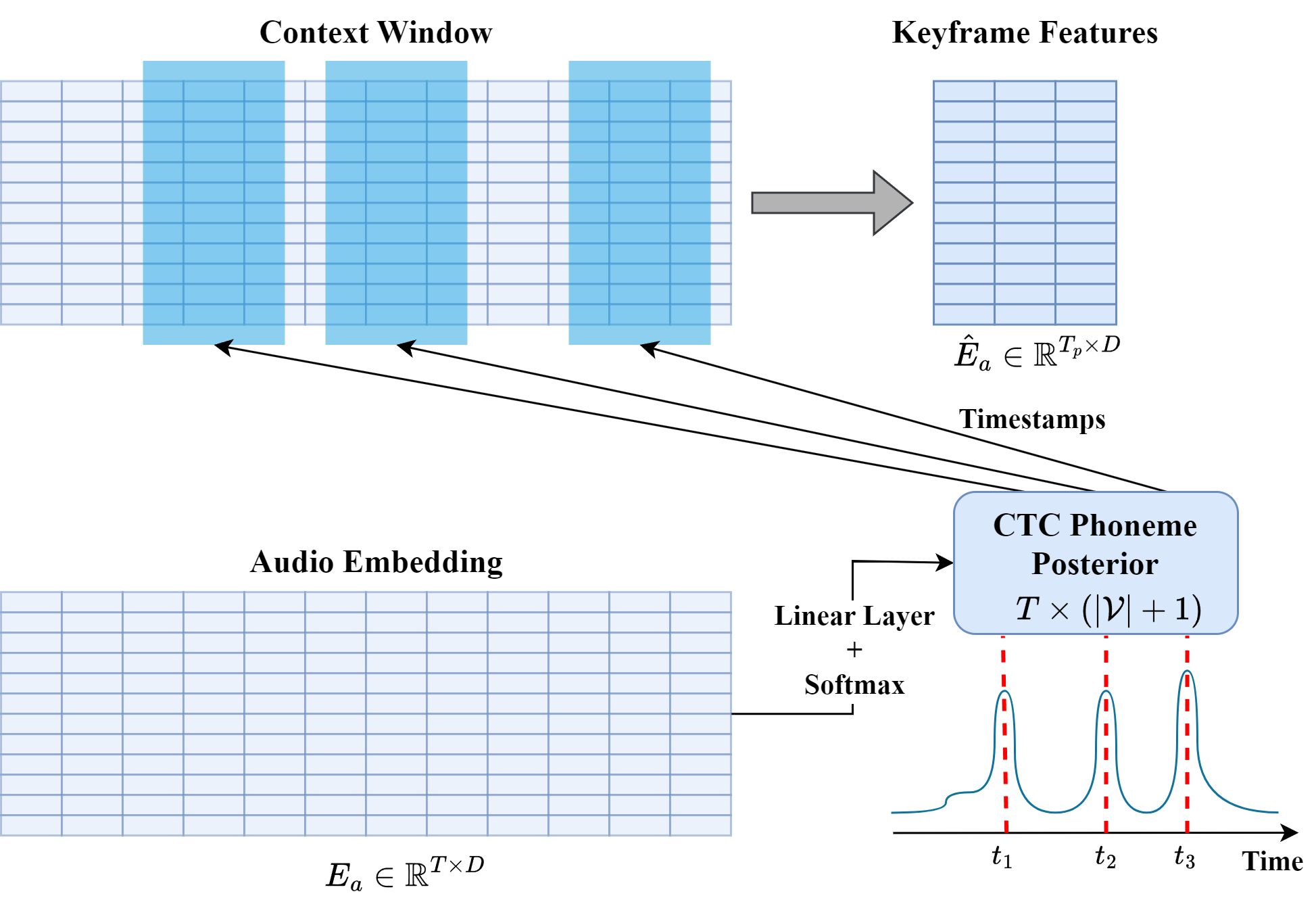}
    \caption{Illustration of the CTC Selector module. The module computes phoneme posterior probabilities to identify key timestamps based on probability peaks. It then extracts and aggregates features from the original audio embeddings within a local context window ($2w+1$) to generate compact, phoneme-aligned keyframe features.}
    \label{fig:ctc}
\end{figure}

\textbf{QbyKeyframe.}
\textit{CTC-Guided Keyframe Selection.}
As illustrated in Fig.~\ref{fig:ctc}, given the audio embedding $E_a \in \mathbb{R}^{T \times D}$
($E_a^q$ or $E_a^s$), we first apply a \textit{linear
projection} followed by softmax to obtain frame-level phoneme posterior
probabilities:
\begin{equation}
\label{eq:ctc_posterior}
P(l_t | E_a) = \text{Softmax}(W_c \cdot E_a + b_c)
\in \mathbb{R}^{T \times (|\mathcal{V}| + 1)},
\end{equation}
where $\mathcal{V}$ is the phoneme vocabulary and the additional
dimension corresponds to the CTC blank token.

\vspace{-2pt}

We then identify keyframes by selecting non-blank peaks from the
CTC posterior. Specifically, for each frame $t$, we obtain the
most likely token $l_t^* = \arg\max_l P(l_t | E_a)$. A frame is
selected as a keyframe if: (1) $l_t^* \neq \texttt{blank}$, and
(2) $l_t^*$ has not been selected by a previous frame (i.e., we
retain only the first occurrence of each unique predicted phoneme).
This \emph{distinct-token} constraint ensures that the selected
keyframes cover diverse phoneme positions rather than
over-representing repeated predictions from CTC's peaky
behavior~\cite{graves2006connectionist}.

To capture local acoustic context around each keyframe, we apply
a symmetric context window of size $2w+1$ centered at each selected
frame position $t_k$, and compute the keyframe representation as
the mean of features within the window:
\begin{equation}
\label{eq:context_window}
\hat{e}_{t_k} = \frac{1}{2w+1} \sum_{j=t_k-w}^{t_k+w} e_j,
\end{equation}
where $e_j$ denotes the $j$-th frame of the projected audio
embedding. The resulting keyframe sequence
$\hat{E}_a \in \mathbb{R}^{T_p \times D}$ has length $T_p$
equal to the number of distinct predicted phonemes, which is
naturally aligned with the phoneme embedding $E_p^s$.

We then compute cosine similarity matrices between $\hat{E}_a^q$ and each of $\hat{E}_a^s$, $E_p^s$, and $E_t^s$, as shown in Equation \ref{eq:cos_sim}:
\begin{equation}
\label{eq:cos_sim}
M_{am} = 
\frac{
    (\hat{\mathbf{E}}_a^q)^\top \mathbf{E}_m^s
}{
    \lVert \hat{\mathbf{E}}_a^q \rVert_F \cdot \lVert \mathbf{E}_m^s \rVert_F
}, 
\quad m \in \{\textit{a}, \textit{p}, \textit{t}\}
\end{equation}
where $a$, $p$, and $t$ denote the audio, phoneme, and text modalities, respectively.

Each similarity matrix $M_{am}$ encodes the frame-level correspondence
between keyframes and enrollment features. Rather than using keyframe
features directly as cross-attention input, we feed the similarity
matrix as queries to attend over the full-utterance representation
$F_m^c$. This design choice is intentional: the similarity matrix
captures \emph{where} keyframes match across modalities, while the
full-utterance vector $F_m^c$ encodes \emph{what} global patterns
exist. Cross-attention between these two complementary views produces
keyframe representations that are both locally precise (from similarity
matching) and globally contextualized (from full-utterance features):
\begin{equation}
\label{eq:cross}
F_m^k = \text{CrossAttn}(Q{=}M_{am},\ K{=}V{=}F_m^c).
\end{equation}
Finally, $F_a^k$, $F_p^k$, and $F_t^k$ are summed to produce the utterance-level posterior probability.

\subsection{Cross-Modal Regularization}
To enhance the robustness of multimodal fusion, we apply
\emph{modality dropout} during training: for each training sample,
we independently zero out each enrollment modality embedding
(audio, phoneme, or text) with probability $p = 0.5$. This
differs fundamentally from temporal masking approaches such as
SpecAugment~\cite{park2019specaugment}, which mask time-frequency
regions within a single modality. Our modality dropout operates
at the \emph{modality level}, forcing the model to make predictions
even when an entire modality stream is absent, thereby encouraging
each modality branch to develop independently informative
representations rather than relying on a single dominant modality.
This strategy is analogous to standard dropout~\cite{srivastava2014dropout}
applied at the modality granularity rather than the neuron level.

\subsection{Learning Strategies}
Each training sample consists of a query audio segment and an enrollment pair (audio + text), where the ground-truth phoneme sequences required for CTC supervision (Eq.~\ref{eq:Lctc}) are obtained via G2P~\cite{g2pE2019}. At inference, only the enrollment text and query audio waveform are needed; the CTC module serves solely as a keyframe selector, requiring no query transcription.

KFC-KWS is trained with a composite loss combining four objectives. Let $\mathrm{BCE}(\ell, y) = -[ y \log \sigma(\ell) + (1{-}y) \log (1{-}\sigma(\ell)) ]$ denote binary cross-entropy, where $\sigma$ is the sigmoid function. The \textbf{utterance-level loss} supervises the fused representation obtained by summing $F_a^k$, $F_p^k$, $F_t^k$ and applying a linear projection to produce logit $\ell_u$:

\begin{equation}
\label{eq:Lutt}
\mathcal{L}_u = \mathrm{BCE}(\ell_u, y_u),
\end{equation}
where $y_u \in \{0,1\}$ is the ground-truth utterance label. The \textbf{sequence-level losses} provide frame-wise supervision at the phoneme and word levels. Frame-level logits $\ell_{m,t}$ are extracted from $\overline{E_m^s}$ via indexing and linear projection:

\begin{equation}
\label{eq:seqbce}
\mathcal{L}_s^m = \frac{1}{N_m} \sum_t \text{mask}_{m,t} \cdot \mathrm{BCE}(\ell_{m,t}, y_{m,t})
\end{equation}
where $\text{mask}_{m,t}$ identifies valid tokens (ignoring padding), $N_m$ is the count of valid elements, and $m \in \{p, t\}$. The \textbf{CTC loss} supervises frame-level phoneme predictions for both query and enrolled audio:

\begin{equation}
\label{eq:Lctc}
\mathcal{L}_c = \mathrm{CTC}(\mathbf{z}^q, \mathbf{p}^q) + \mathrm{CTC}(\mathbf{z}^s, \mathbf{p}^s),
\end{equation}
where $\mathbf{z}^{q/s}$ are linear projections of the audio features and $\mathbf{p}^{q/s}$ the corresponding ground-truth phoneme sequences. The total training objective is:

\begin{equation}
\label{eq:loss}
\mathcal{L}_{\text{total}} = \mathcal{L}_u + \mathcal{L}_s^p + \mathcal{L}_s^t + \lambda \mathcal{L}_c,
\end{equation}
where $\lambda$ balances the CTC loss against the other terms.

\section{Experiments}
\label{sec:exp}

\subsection{Datasets}
We evaluate KFC-KWS on LibriPhrase\footnote{\url{https://github.com/gusrud1103/LibriPhrase.git}},
a benchmark derived from LibriSpeech~\cite{panayotov2015librispeech}
consisting of short phrases (1--4 words). The training set is extracted from the train-clean-100 and train-clean-360 subsets, while the evaluation set is sourced from the train-other-500 subset. Based on the phonetic similarity between positive and negative pairs, the evaluation set is
divided into LibriPhrase-Easy (LPE) and LibriPhrase-Hard (LPH). The LPH subset contains a large number of phonetically confusable keyword pairs, making it particularly challenging.
\subsection{Implementation Details}
For the pre-trained encoders, we use XLS-R
(0.3B)~\cite{babu2022xls} for audio encoding,
G2P~\cite{g2pE2019, lee2024iphonmatchnet} to convert text into 64-dimensional phoneme embeddings, and multilingual
DistilBERT~\cite{sanh2019distilbert} for text encoding.
The query and enrolled audio share the same XLS-R encoder with
frozen weights. The context window size $w$ for keyframe
extraction is set to 2 (i.e., a window of 5 frames).
Projection layers map all modality features to a 128-dimensional
shared space. The QbyOmni module uses a 2-layer Transformer
encoder with feed-forward dimension 512. The GRU has a hidden
size of 64. The CTC loss weight $\lambda = 0.2$ and modality
dropout rate $p = 0.5$.
The total number of trainable parameters in KFC-KWS is
approximately \textbf{2.0 M} (excluding frozen pre-trained
encoders). All experiments are conducted on a single NVIDIA 4080
Super GPU with batch size 512 and Adam optimizer
(lr = 0.001) for 50 epochs.

\section{Results}
\label{sec:res}
\subsection{Overall Performance}
We report both subset-specific and \emph{balanced} metrics
(arithmetic mean of LPH and LPE, denoted ``Bal.'' in the tables), as practical deployment
involves both easy and hard keywords.

\subsubsection{Comparison with State-of-the-Art Methods}
As shown in Table~\ref{tab:without_aug}, KFC-KWS achieves the highest balanced AUC (98.06\%) among all methods, surpassing the runner-up HyperSpotter-c (97.98\%) while using only \textbf{2.0\,M} trainable parameters---2.75$\times$ fewer than HyperSpotter-c (5.5\,M). On the challenging LPH subset, KFC-KWS attains the best AUC (96.54\%) and EER (9.13\%). Notably, it outperforms the strong multimodal baseline PLCL by 0.98\% in AUC and 0.83\% in EER, and surpasses HyperSpotter-c by 0.47\% in AUC, demonstrating that CTC-guided keyframe selection effectively isolates the phoneme positions where confusable keywords differ. Regarding the balanced EER, KFC-KWS (5.68\%) remains competitive with the top-performing baselines DS-KWS-M1 (5.27\%) and PLCL (5.59\%). The gap is primarily driven by the LPE trade-off (2.22\% vs.\ 0.52\% for DS-KWS-M1), where full-sequence models excel at easy samples, while our method prioritizes the discriminability of hard samples.

\begin{table}[htbp]
\centering
\footnotesize
\setlength{\tabcolsep}{4pt}
\renewcommand{\arraystretch}{1.1}
\caption{Performance on LibriPhrase \textbf{without augmentation}. Best in \textbf{bold}, second best \underline{underlined}. \textbf{Bal.}\,=\,(LPH\,+\,LPE)/2.}
\label{tab:without_aug}
\resizebox{\columnwidth}{!}{
\begin{tabular}{l c cc cc cc}
\toprule
\multirow{2}{*}{\textbf{Method}} & \multirow{2}{*}{\textbf{Params}}
& \multicolumn{2}{c}{\textbf{AUC (\%)$\uparrow$}}
& \multicolumn{2}{c}{\textbf{EER (\%)$\downarrow$}}
& \multicolumn{2}{c}{\textbf{Bal.}} \\
\cmidrule(lr){3-4} \cmidrule(lr){5-6} \cmidrule(lr){7-8}
& & LPH & LPE & LPH & LPE & AUC$\uparrow$ & EER$\downarrow$ \\
\midrule
EMKWS \cite{nishu2023matching} & 3.7M
  & 73.58 & 97.83 & 32.9 & 8.42 & 85.71 & 20.66 \\
iPhonMatchNet \cite{lee2024iphonmatchnet} & 0.7M
  & 88.23 & 99.59 & 19.70 & 2.40 & 93.91 & 11.05 \\
CED \cite{nishu2024flexible} & 3.8M
  & 89.20 & 99.94 & 18.40 & 0.80 & 94.57 & 9.60 \\
HyperSpotter-c(4) \cite{segal2025keyword} & 5.5M
  & \underline{96.07} & 99.89 & 10.45 & 1.08 & \underline{97.98} & 5.77 \\
SLiCK \cite{nishu2025slick} & 0.6M
  & 94.90 & 99.82 & 11.10 & 1.78 & 97.36 & 6.44 \\
MM-KWS \cite{ai2024mm} & 3.9M
  & 94.02 & \textbf{99.98} & 12.46 & \textbf{0.41} & 97.00 & 6.44 \\
PLCL \cite{kewei2024phoneme} & 40.0M
  & 95.56 & \underline{99.95} & \underline{9.96} & 1.21 & 97.76 & \underline{5.59} \\
DS-KWS-M1 \cite{ai2025dual} & 4.1M
  & 95.77 & \textbf{99.98} & 10.02 & \underline{0.52} & 97.88 & \textbf{5.27} \\
\midrule
\textbf{KFC-KWS} & 2.0M
  & \textbf{96.54} & 99.58 & \textbf{9.13} & 2.22 & \textbf{98.06} & 5.68 \\
\bottomrule
\end{tabular}
}
\end{table}
\vspace{-2pt}

As shown in Table~\ref{tab:with_aug}, when equipped with augmentation strategies (modality dropout for KFC-KWS), our method achieves the best balanced AUC across all configurations (\textbf{98.73\%}). Specifically on LPH, it reaches 97.65\% AUC and 7.75\% EER, outperforming the strongest augmented baseline PLCL$^\dagger$ by margins of 1.06\% and 0.72\%, respectively. The balanced EER (4.85\%) is highly competitive, approaching that of PLCL$^\dagger$ (4.52\%), confirming that KFC-KWS provides a robust solution for realistic deployment scenarios containing both easy and hard keywords.

\begin{table}[htbp]
\centering
\footnotesize
\setlength{\tabcolsep}{4pt}
\renewcommand{\arraystretch}{1.1}
\caption{Performance on LibriPhrase \textbf{with augmentation}$^\dagger$. Notations follow Table~\ref{tab:without_aug}.}
\label{tab:with_aug}
\resizebox{\columnwidth}{!}{
\begin{tabular}{l cc cc cc cc}
\toprule
\multirow{2}{*}{\textbf{Method$^\dagger$}}
& \multicolumn{2}{c}{\textbf{AUC (\%)$\uparrow$}}
& \multicolumn{2}{c}{\textbf{EER (\%)$\downarrow$}}
& \multicolumn{2}{c}{\textbf{Bal.}}
& \multicolumn{2}{c}{\textbf{$\Delta$\,Bal.}} \\
\cmidrule(lr){2-3}
\cmidrule(lr){4-5}
\cmidrule(lr){6-7}
\cmidrule(lr){8-9}
& LPH & LPE & LPH & LPE & AUC$\uparrow$ & EER$\downarrow$ & AUC & EER \\
\midrule
CED$^\dagger$ \cite{nishu2024flexible}
  & 92.70 & 99.84 & 14.40 & 1.70 & 96.27 & 8.05
  & +1.70 & $-$1.55 \\
AdaKWS-Small$^\dagger$ \cite{navon2024open}
  & 95.09 & 99.82 & 11.48 & 1.21 & 97.46 & 6.35
  & -- & -- \\
MM-KWS$^\dagger$ \cite{ai2024mm}
  & 96.25 & \underline{99.95} & 9.30 & \underline{0.68} & 98.10 & 4.99
  & +1.10 & $-$1.45 \\
PLCL$^\dagger$ \cite{kewei2024phoneme}
  & \underline{96.59} & \textbf{99.97} & \underline{8.47} & \textbf{0.57} & \underline{98.28} & \textbf{4.52}
  & +0.52 & $-$1.07 \\
\midrule
\textbf{KFC-KWS$^\dagger$}
  & \textbf{97.65} & 99.81 & \textbf{7.75} & 1.94 & \textbf{98.73} & \underline{4.85}
  & +0.67 & $-$0.83 \\
\bottomrule
\end{tabular}
}
{\raggedright\scriptsize
$\dagger$: modality dropout for KFC-KWS; method-specific strategies for others. ``--'': not reported. $\Delta$\,Bal.: gain over Table~\ref{tab:without_aug}.\par}
\end{table}
\vspace{-2pt}

\subsubsection{Effect of Augmentation \& Trade-off Analysis}
The $\Delta$\,Bal.\ columns in Table~\ref{tab:with_aug} reveal that methods like CED and MM-KWS benefit significantly from augmentation (+1.70\% and +1.10\% AUC), indicating that data diversity helps mitigate their reliance on global context. KFC-KWS achieves a solid +0.67\% gain from modality dropout alone, showing that the keyframe architecture is inherently robust but still responsive to regularization.
Regarding the LPH--LPE trade-off, a consistent pattern emerges: our method sacrifices a small degree of performance on easy samples (LPE EER $\approx$ 2\%) to achieve substantial gains on hard samples (LPH EER $\approx$ 7--9\%). Since LPE samples are acoustically distinct, full-sequence methods achieve near-perfect metrics (EER $<$ 1\%), whereas our sparse keyframe selection naturally discards some redundant cues beneficial for these easy cases. However, the superior balanced metrics confirm that the significant improvements on confusable keywords (LPH) far outweigh the marginal costs on easy ones.

\subsection{Ablation Studies}
Table~\ref{tab:ablation} shows the contribution of each modality
encoder. Removing the phoneme encoder causes the largest
degradation (5.75\% AUC drop on LPH), confirming that phoneme-level
information is essential for the CTC-guided keyframe strategy.
Removing the text encoder yields an interesting pattern: LPH
performance degrades moderately while LPE improves to 99.95\% AUC,
suggesting that text-level semantic features primarily benefit
hard-case discrimination by providing additional word-level cues
that complement acoustic similarity. Removing the audio encoder
shows balanced degradation across both subsets, indicating that
audio features contribute consistently to both matching scenarios.

\begin{table}[htbp]
\centering
\footnotesize
\setlength{\tabcolsep}{5pt}
\renewcommand{\arraystretch}{1.1}
\caption{Ablation studies of KFC-KWS. ``w/o'' denotes removal of the corresponding pre-trained encoder during enrollment. Best in \textbf{bold}, second best \underline{underlined}.}
\label{tab:ablation}
\begin{tabular}{lcccccc}
\toprule
\multirow{2}{*}{\textbf{Method}} 
& \multicolumn{2}{c}{\textbf{AUC (\%)$\uparrow$}} 
& \multicolumn{2}{c}{\textbf{EER (\%)$\downarrow$}}
& \multicolumn{2}{c}{\textbf{Bal. (\%)}} \\
\cmidrule(lr){2-3} \cmidrule(lr){4-5} \cmidrule(lr){6-7}
& \textbf{LPH} & \textbf{LPE} & \textbf{LPH} & \textbf{LPE} & \textbf{AUC}$\uparrow$ & \textbf{EER}$\downarrow$ \\
\midrule
KFC-KWS\textsuperscript{\dag}
& \textbf{97.65} & \underline{99.81}
& \textbf{7.75}  & \underline{1.94}
& \textbf{98.73} & \textbf{4.85} \\

w/o audio
& 96.78 & 99.07
& \underline{9.18}  & 4.86
& 97.93 & \underline{7.02} \\

w/o text
& \underline{97.33} & \textbf{99.95}
& 18.90 & \textbf{0.77}
& \underline{98.64} & 9.84 \\

w/o phoneme
& 91.90 & 97.52
& 18.32 & 8.88
& 94.71 & 13.60 \\
\bottomrule
\end{tabular}
\end{table}

\section{Conclusion}
\label{sec:cnl}
We presented KFC-KWS, a multimodal framework for user-defined keyword
spotting that leverages CTC-guided keyframe selection to achieve
fine-grained phoneme-level cross-modal matching. By exploiting the
peaky posterior distributions of CTC, our approach identifies
the most phonetically salient frames and fuses them with full-utterance
representations, enabling precise discrimination of phonetically
confusable keywords. Experiments on LibriPhrase demonstrate that
KFC-KWS achieves the best balanced performance across both hard and
easy evaluation settings without requiring complex data augmentation
or external resources. Future work will explore adaptive keyframe
selection strategies and evaluation on noise-robust benchmarks
for real-world deployment.

\newpage
\section{Acknowledgments}
This work was supported in part by the Yangtze River Delta Science and Technology Innovation Community Joint Research Project 2024CSJGG1100 and in part by the Key R\&D Program Project of Zhejiang Province (2025C01167). 
The authors would like to express their sincere gratitude to \textbf{Yu Xi} from Shanghai Jiao Tong University for his valuable guidance and constructive comments, which greatly benefited this research.

\section{Use of Generative AI Disclosure}
Generative AI tools were used for editing and polishing the
manuscript text. All experimental results, method design, and
scientific conclusions are the sole responsibility of the authors.

\bibliographystyle{IEEEtran}
\bibliography{mybib}

\end{document}